\documentclass[final,5p,times,twocolumn,authoryear]{elsarticle}
\usepackage{amssymb}
\usepackage{lipsum}
\usepackage{blindtext}
\usepackage{hyperref}
\usepackage{amsmath,amssymb}
\usepackage{float}
\usepackage{microtype}
\usepackage{ragged2e}  

\usepackage{graphicx}
\usepackage{bm}
\usepackage{latexsym}
\usepackage{epsfig}
\usepackage{psfrag}
\usepackage[dvipsnames]{xcolor}
\usepackage{subfigure}
\usepackage{subcaption}
\usepackage{graphicx}
\usepackage{multirow}
\usepackage{amssymb}
\usepackage{amsmath}
\usepackage{bm}
\usepackage{latexsym}
\usepackage{epsfig}
\usepackage{psfrag}
\usepackage[normalem]{ulem}
\usepackage{textcomp}
\usepackage{color}
\usepackage{pstricks}
\usepackage[utf8]{inputenc}
\usepackage{comment}
\usepackage{hyperref}
\usepackage[mathlines]{lineno}
\usepackage{hyperref}
\usepackage[all]{hypcap}
\hypersetup{  
    colorlinks=true,
    linkcolor=blue,
    filecolor=red,      
    urlcolor=cyan,
    citecolor=green,
    }

\def\beq{\begin{equation}}
\def\eeq{\end{equation}}
\def\bea{\begin{eqnarray}}
\def\eea{\end{eqnarray}}
\def\be{\begin{equation}}
\def\ee{\end{equation}}

\def\bse{\begin{subequations}}
\def\ese{\end{subequations}}

\def\ee{\eta_{\rm e}}

\def\Mpl{M_{_\mathrm{P}}}

\def\d{\mathrm{d}}

\def\w{w_{\rm RH}}

\journal{Physics Letters B}

\begin{document}

\begin{frontmatter}

\title{Improved Predictions on Higgs-Starobinsky Inflation and Reheating with ACT DR6 and Primordial Gravitational Waves}

\author[first]{Md Riajul Haque}
\author[first]{Sourav Pal}
\author[first]{Debarun Paul}
\affiliation[first]{organization={Physics and Applied Mathematics Unit},
            addressline={Indian Statistical Institute}, 
            city={203 B.T. Road},
            postcode={Kolkata 700108},
            country={India}}

\begin{abstract}
In this letter, we investigate the implications of recent CMB observations for Higgs–Starobinsky inflationary models and their associated reheating dynamics, utilising data from ACT DR6, Planck 2018, BICEP/Keck 2018, and DESI, collectively referred to as P-ACT-LB-BK18. In addition to direct CMB constraints, we incorporate indirect bounds arising from the potential overproduction of primordial gravitational waves (PGWs), particularly through limits on the effective number of relativistic species, $\Delta N_{\rm eff}$, during Big Bang Nucleosynthesis (BBN). These constraints become especially relevant in scenarios featuring a stiff post-inflationary equation of state $w_{\rm RH}\geq 0.58$. Our analysis shows that, when both P-ACT-LB-BK18 data and $\Delta N_{\rm eff}$ bounds are considered, the viable number of inflationary e-folds is restricted to the range $57.9$–$62.2$ at the $2\sigma$ confidence level (C.L.). Correspondingly, the reheating temperature is constrained to lie between the BBN energy scale and $10^{12}$ GeV, with the post-inflationary equation-of-state parameter satisfying $w_{\rm RH} > 0.41$. However, no parameter space remains viable at the $1\sigma$ C.L. once $\Delta N_{\rm eff}$ constraints from PGWs are included, rendering the Higgs–Starobinsky model highly restricted.
\end{abstract}

\end{frontmatter}

\section{Introduction}
\label{sec:intro}

In recent years, our picture of the early Universe has been transformed by a new era of precision cosmology. Measurements of the cosmic microwave background (CMB), particularly from Planck18~\cite{Planck:2018jri,Planck:2018vyg}, BICEP/$Keck$~\cite{BICEP2:2018kqh,BICEP:2021xfz}, and the Atacama Cosmology Telescope (ACT)~\cite{ACT:2025fju,ACT:2025tim}, have placed stringent limits on inflationary theories. These observations have not only improved the accuracy of the observables, \textit{i.e.} scalar spectral index $n_s$ and the tensor-to-scalar ratio $r$, but also provided deeper insight into the physics driving inflation and post-inflationary reheating phase~\cite{Kolb:1990vq,Shtanov:1994ce,Kofman:1997yn,Allahverdi:2010xz,Lozanov:2019jxc,Barbieri:2025moq}. 

Among many inflationary models proposed over the decade (see Ref.~\cite{Martin:2013tda,Odintsov:2023weg} for all proposed models), Starobinsky is the simplest and theoretically well-motivated model for cosmic inflation~\cite{Starobinsky:1980te,Kallosh:2013hoa,Ellis:2013nxa,Kallosh:2013yoa}. 
The Higgs inflation \cite{Bezrukov:2007ep} model offers a different route to inflation by identifying the Standard Model Higgs boson as the inflaton. In the Einstein frame, both converges to a potential of the same form, which is a special case of more general $\alpha$-attractor E model \cite{Kallosh:2013yoa,Kallosh:2013hoa,Roest:2013fha,Ferrara:2013rsa,Kallosh:2013tua,Cecotti:2014ipa,Kallosh:2014rga,Galante:2014ifa}. 

The improvements in observational precision have culminated most recently in the Data Release 6 (DR6) from the ACT, which has provided improved measurements of the CMB power spectrum at high multipoles. When combined with Planck18 (with lensing) and baryon acoustic oscillation (BAO) measurements from DESI, the ACT data (P-ACT-LB) tighten the constraint on $n_s$ to $0.9743 \pm 0.0034$, approximately $2\sigma$ deviation from the value inferred using Planck only measurement. 
Driven by these advancements, a range of inflationary models, for instance warm inflation~\cite{Berera:1995ie,Bartrum:2013fia,Berera:2023liv}, Higgs inflation~\cite{Bezrukov:2007ep}, have been re-examined in light of the latest ACT data~\cite{Kallosh:2025rni,Aoki:2025wld,Berera:2025vsu,Dioguardi:2025vci,Gialamas:2025kef,Salvio:2025izr,Antoniadis:2025pfa,Kim:2025dyi,Dioguardi:2025mpp,Gao:2025onc,He:2025bli,Pallis:2025epn,Drees:2025ngb,Gialamas:2025ofz,Yin:2025rrs}. 
Until the recent ACT data release, Higgs-Starobinsky model was always favoured by observations~\cite{Martin:2013tda,Planck:2018jri}. However, recent analysis using ACT data~\cite{ACT:2025tim} has shown that the Higgs–Starobinsky model is excluded at the $2\sigma$ confidence level. Subsequent studies~\cite{Drees:2025ngb,Zharov:2025evb,Liu:2025qca} have demonstrated that, by incorporating a post-inflationary reheating phase—defined as the transition from the end of inflation to the onset of the radiation-dominated era—with a very stiff equation of state (EoS), the model can still be made consistent with observations at the $1\sigma$ C.L.

All these aforementioned analyses incorporate BBN \textit{only} limit on reheating temperature, \textit{without} accounting any impacts from primordial gravitational waves (PGWs) overproduction.
One of the key prediction of the inflation is the generation of tensor fluctuations at the early Universe which, in turn, produce 
PGWs~\cite{Grishchuk:1974ny,Starobinsky:1979ty,Boyle:2005se,Watanabe:2006qe,Saikawa:2018rcs,Caprini:2018mtu}. 
Thus, the post-inflationary reheating phase has a strong influence on the spectrum of these PGWs, especially for the modes which re-enter the horizon during reheating ~\cite{Bernal:2019lpc,Haque:2021dha,Maity:2024odg,Maity:2024cpq,Ghoshal:2024gai,Barman:2023ktz}. 
If the effective EoS during reheating is stiff ($w_{\rm RH}>1/3$), the resulting gravitational wave spectrum becomes blue-tilted, leading to an enhanced amplitude at high frequencies. Such an enhancement increases the total energy density in PGWs, which contributes to the effective number of relativistic species, $\Delta N_{\rm eff}$, a quantity constrained by observations of both Big Bang Nucleosynthesis (BBN) and the cosmic microwave background (CMB). 
As a result, current bounds on $\Delta N_{\rm eff}$ provide an indirect but powerful handle on reheating dynamics, enabling lower limits to be placed on the reheating temperature $T_{\rm RH}$ in scenarios with stiff post-inflationary expansion.
The analysis of our work, thus, completes the standing of Higgs-Starobinsky model in the context of recent P-ACT-LB along with BK18 observations (P-ACT-LB-BK18), accounting finite duration of reheating dynamics as well as the constraints on PGWs overproduction, that was not considered in the earlier works~\cite{Drees:2025ngb,Zharov:2025evb,Liu:2025qca}.

The paper is organized as follows: In Sec.~~\ref{sec:inflation_overview}, we outline the Higgs–Starobinsky inflationary model, reheating dynamics, and the constraints on PGWs from the $\Delta N_{\rm eff}$ bound at BBN. In Sec.~~\ref{sec:results}, we present our results, including constraints from both reheating dynamics and PGW overproduction. Finally, Sec.~\ref{sec:conclusion} summarizes our findings and outlines directions for future research.

\section{Model}
\label{sec:inflation_overview}

\subsection{Higgs-Starobinsky Inflation}
\label{subsec:HS_inflation}
 Both the Starobinsky and Higgs inflationary models arise from distinct theoretical frameworks, yet they exhibit a deep equivalence when analyzed in the Einstein frame. Starobinsky inflation originates from a higher-order modification of general relativity, where a term quadratic in the Ricci scalar is added to the gravitational action. In contrast, Higgs inflation extends the Standard Model by introducing a non-minimal coupling between the Higgs field and the spacetime curvature. 

Despite these different starting points, both models can be reformulated through a conformal transformation followed by field redefinition. In the resulting Einstein frame, each theory becomes a standard scalar field model minimally coupled to gravity, with identical potential shapes for the inflaton.
The scalar potential common to both models takes the form: 
\begin{equation}\label{eq:HS_potential}
V(\phi) = \beta \left( 1 - e^{-\sqrt{\frac{2}{3}} \frac{\phi}{M_{\rm P} }}\right)^2 \, ,   
\end{equation}
where $\phi$ is the canonically normalized scalar field and $M_{\rm P}$ is the reduced Planck mass. For Starobinsky inflation, $\beta_S=1/(4\alpha)$, with $\alpha \equiv \frac{M_{\rm P}^2}{12 M^2}$. For Higgs inflation, $\beta_H = \lambda M_{\rm P}^4/\xi^2$, where $\lambda$ is the Higgs self-coupling and $\xi$ is the non-minimal coupling to gravity. Here, prefixes $S$ and $H$ stand for Starobinsky and Higgs model respectively. Including the aforementioned coupling, the non-canonical Lagrangians in Jordan frame read as,
\begin{align}
\mathcal{L}_S &= \frac{M_{\rm P}^2}{2} R \left(1 + \alpha R \right) + \cdots \label{eq:LS} \\
\mathcal{L}_H &= \frac{M_{\rm P}^2}{2} R + \frac{2\xi R}{M_{\rm P}^2} h^2 - \frac{1}{2} \partial_\mu h \partial^\mu h - \frac{\lambda}{4} h^4 + \cdots \label{eq:LH}
\end{align}

During Higgs inflation, one should note that $\xi>1, h/M_{\rm P}>1$. Once the theories are transformed to the Einstein frame, the inflaton fields relate to the original variables logarithmically: \\
(i) In the Starobinsky case, $\phi=\sqrt{2/3 } \, M_{\rm P} \ln{(1+2\alpha R)}$, (ii) In the Higgs case, $\phi=\sqrt{2/3} \, M_{\rm P} \ln{(1+\xi h^2/M_{\rm P}^2)}$. This similarity in the Einstein frame implies that both models, once the inflationary energy scale is fixed to match the CMB normalization, lead to nearly identical predictions for observable quantities such as the scalar spectral index $n_s$ and the tensor-to-scalar ratio $r$.
Assuming slow-roll approximations, the dynamics of the inflaton field is primarily governed by the inflationary potential, which can be expressed in terms of $r$ and the amplitude of the scalar power spectra $A_{s}$ as,
\begin{eqnarray}
H_{\rm k} &=& \frac{\pi M_{\rm P}\sqrt{r\,A_s}} {\sqrt{2}} \simeq  \sqrt{\frac{V(\phi_{\rm k})}{3 M_{\rm P}^2} } \, ,\label{eq:Hk} 
\end{eqnarray} 
where $H_k$ is the Hubble parameter at the point of Horizon crossing of the CMB scale.
Inflation's duration is typically quantified by the total number of e-folds, $N_k$, spanning from the horizon exit of a CMB perturbation with comoving wavenumber $k$ to the end of the inflationary phase. For Higgs-Starobinsky model, $N_k$ is determined by,
\begin{eqnarray}\label{eq:Nk}
    N_k =\frac{3 }{4}\left(\exp{\sqrt{\frac{2}{3}}\frac{\phi_{k}}{M_{\rm P}}}-\exp{\sqrt{\frac{2}{3}}\frac{\phi_{\rm end}}{M_{\rm P}}} \right) \nonumber \\ 
    -\sqrt{\frac{3}{8  M_{\rm P}^2}}(\phi_k - \phi_{\rm end})\,,
\end{eqnarray}
where $\phi_k$ and $\phi_{\rm end}$ are field values at the time of horizon crossing of CMB scale and at the end of inflation respectively ~\cite{Cook:2015vqa}.
In the following, we provide a brief overview of reheating dynamics.

\begin{figure*}
\centering
\includegraphics[width=0015.0cm,height=10.1cm]{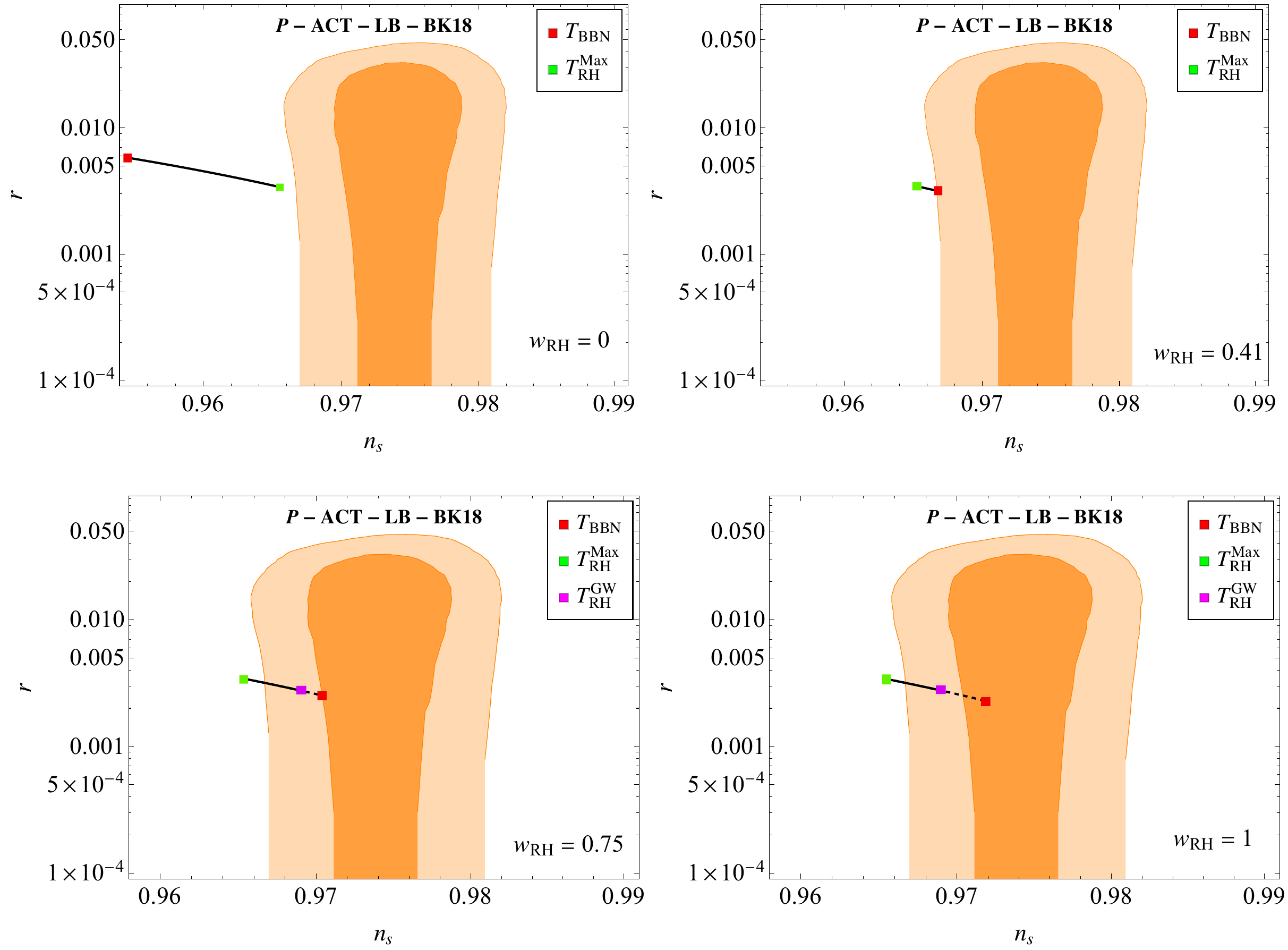}
\caption[]{\justifying \em Predictions of Higgs-Starobinsky inflation for a range of post-inflationary equations of state are presented in the $(n_s, r)$ plane, overlaid with the latest combined observational constraints from \text{P-ACT-LB-BK18}. The dark and light orange regions denote the $1\sigma$ (68\% C.L.) and $2\sigma$ (95\% C.L.) confidence intervals, respectively. The reheating temperature is varied between $T_{\rm BBN}$ and the maximum reheating temperature $T_{\rm RH}^{\rm Max}$. A critical temperature scale, $T_{\rm RH}^{\rm GW}$, is highlighted with a magenta square, corresponding to the upper bound from $\Delta N_{\rm eff}$ constraints due to the overproduction of primordial gravitational waves.
}
\label{fig:nsr_model}
\end{figure*}

\subsection{Reheating Dynamics}
\label{subsec:reheating}

In minimal extensions of the Standard Model, the radiation-dominated (RD) era is preceded by a reheating phase following inflation~~\cite{Kolb:1990vq,Shtanov:1994ce,Kofman:1997yn,Allahverdi:2010xz,Lozanov:2019jxc}. This epoch is characterized by the reheating temperature $T_{\rm RH}$ and an effective equation-of-state parameter $w_{\rm RH}$, defined as the ratio of pressure to energy density. Reheating ends when the energy densities of the inflaton and radiation become comparable, $\rho_\phi \simeq \rho_R$, marking the onset of the RD era. Assuming a constant $w_{\rm RH}$, the energy density at the end of reheating (i.e., $\rho_{\rm RH}$) can be related to that at the end of inflation (i.e., $\rho_{\rm end}$) as
\begin{eqnarray}
\frac{\rho_{\rm end}}{\rho_{\rm RH}} = \left( \frac{a_{\rm end}}{a_{\rm RH}} \right)^{-3(1 + w_{\rm RH})}.
\end{eqnarray}
where $a_{\rm end}$ and $a_{\rm RH}$ denote the scale factor at the end of inflation and reheating, respectively. The energy density at the end of reheating is given by $\rho_{\rm RH} = \frac{\pi^2}{30} g_{\rm \ast RH}\, T_{\rm RH}^4$, allowing the reheating temperature to be expressed in terms of the inflationary energy scale and equation of state as
\begin{eqnarray}
\label{eq:trh_w_nrh}
T_{\rm RH} \simeq \left(\frac{90 \Mpl^2 H_{\rm end}^2}{\pi^2 g_{\rm \ast RH}}\right)^{1/4} e^{-\frac{3}{4}N_{\rm RH}(1+w_{\rm RH})},
\end{eqnarray}
where $N_{\rm RH}$ is the number of e-folds between the end of inflation and the end of reheating, and $g_{\rm \ast RH}$ denotes the number of relativistic degrees of freedom in the thermal bath at that time.

Assuming conservation of comoving entropy from the end of reheating to the present, the reheating temperature can also be related to the current CMB temperature ($T_0 = 2.735 \, {\rm K}$) via~\cite{Dai:2014jja, Cook:2015vqa}
\begin{equation}
\label{eq:trh_w_nrh_ne}
T_{\rm RH} = \left(\frac{43}{11 g_{*S,\rm RH}}\right)^{1/3} T_0 \frac{H_{\rm k}}{k_{\ast}} e^{-(N_{k} + N_{\rm RH})},
\end{equation}
where $N_{k}$ is the number of e-folds between the CMB pivot scale $k_{\ast}$ leaving the horizon and the end of inflation.

Equating Eqs.~~\eqref{eq:trh_w_nrh} and~~\eqref{eq:trh_w_nrh_ne}, one obtains an expression for $N_k$ in terms of the reheating temperature:
\begin{align}
\label{eq:Ninf_reheating}
    N_{k} = \,&\log \left[
\left(\frac{43}{11g_{\rm \ast S, RH}}\right)^{1/3} T_0 \frac{H_{\rm k}}{k_{\ast}}  \right . \nonumber \\
& \times \left. T_{\rm RH}^{\frac{4}{3(1 + \w)} - 1} 
\left(\frac{\pi^2g_{\rm \ast RH}}{90 \Mpl^2 H_{\rm end}^2}\right)^{\frac{1}{3(1 + \w)}}
\right]
\end{align}
By comparing the above expression with the model-specific relation given in Eq.~\eqref{eq:Nk}, one can establish a direct connection between the inflationary model parameters and the subsequent post-inflationary dynamics. In the following, we turn to another important constraint arising from the overproduction of PGWs. These GWs originate from vacuum tensor fluctuations during inflation and evolve through the post-inflationary universe. Their contribution to the effective number of relativistic species, $\Delta N_{\rm eff}$, can become significant—particularly in scenarios with a very stiff post-inflationary EoS ($w_{\rm RH}$). Such contributions must be carefully accounted for, as they impose stringent bounds on the reheating history and the viability of the inflationary model.


\subsection{$\Delta N_{\rm eff}$ bound on PGWs}
\label{subsec:PGW}
 Additional light degrees of freedom in the early universe, quantified by the excess effective number of relativistic species $\Delta N_{\rm eff}$, influence the expansion rate during BBN, affecting light element abundances and constraining new physics. Primordial gravitational waves, particularly the primary one arising from tensor perturbations during inflation, which cannot be neglected, contribute to $\Delta N_{\rm eff}$ if their energy density remains significant. This is especially relevant when a post-inflationary phase with equation-of-state $w_{\rm RH} > 1/3$ enhances the high-frequency tail of the PGW spectrum prior to radiation domination. Recent measurements from ACT together with Planck constraining $\Delta N_{\rm eff} \leq 0.17$ at 95\% C.L \cite{ACT:2025fju,ACT:2025tim}. In line with \cite{Chakraborty:2023ocr,Maity:2024odg}, this bound translates into the following integral constraint on the present-day energy density of PGWs:

\begin{align}
\int_{k_{\rm RH}}^{k_{\rm end}}\frac{\d k}{k}\,\Omega^{\rm (0)}_{\rm GW}(k)\,h^2
\leq \frac{7}{8}\,\left(\frac{4}{11}\right)^{4/3}\,\Omega^{\rm (0)}_{\gamma}\,h^2\,\Delta N_\mathrm{eff},
\label{eq:deltaneff}
\end{align}
where $\Omega^{\rm (0)}{\gamma},h^2\simeq 2.47\times10^{-5}$ denotes the current photon energy density. This bound becomes particularly significant in scenarios where  $w_{\rm RH}>1/3$, as PGWs exhibit a growing spectral energy density for modes re-entering the horizon during reheating, i.e., $k > k_{\rm RH}$ (see~\cite{Chakraborty:2023ocr,Maity:2024odg,Haque:2025uri} for more details).

\begin{table*}[!ht]
    \centering
    \renewcommand{\arraystretch}{1.2}
    \begin{tabular}{c || c | c | c | c | c}
        \hline\hline
        $w_{\rm RH}$ & 0.58 & 0.7 & 0.8 & 0.9 & 1.0 \\[0.5ex]
        \hline
        $T_{\rm RH}^{\rm GW}$ (GeV) 
        & $4.0\times 10^{-3}$ 
        & $1.8\times 10^2$ 
        & $2.6\times 10^4$ 
        & $6.5\times 10^5$ 
        & $6.0\times 10^6$ \\[1ex]
        \hline\hline
    \end{tabular}
    \caption{Numerical values of $T_{\rm RH}^{\rm GW}$ for different values of $w_{\rm RH}$.}
    \label{GWtre}
\end{table*}

Utilizing the form of $\Omega^{\rm (0)}_{\rm GW}$ in RD i.e, $\Omega^{\rm (0)}_{R}$ , the above inequality can be rewritten as:

\begin{eqnarray}
\Omega^{\rm (0)}_{R}\,h^2\,\frac{H_{\rm end}^2}{12\, \pi^2\, M_{\rm P}^2}\, \frac{\mu(w_{\rm RH})\,(1+3\,w_{\rm RH})}{2\,\pi\,
(3\,w_{\rm RH} -1)}\,  \left(\frac{k_{\rm end}}{k_{\rm RH}}\right)^\frac{6\,w_{\rm RH} -2}{1+3\,w_{\rm RH}}  \nonumber \\ 
\leq 5.61\times 10^{-6}\,\Delta N_{\rm eff} \, .\label{Eq:BBNapprox}
\end{eqnarray}
Following~\cite{Chakraborty:2023ocr,Haque:2021dha,Maity:2024odg}, once the ratio of wavenumbers corresponding to horizon re-entry at the end of inflation and at the end of reheating is determined, one can derive a lower bound on the reheating temperature
$T_{\rm RH}$ by combining Eqs.~\eqref{eq:deltaneff} and \eqref{Eq:BBNapprox}. This follows as
\begin{eqnarray}
T_{\rm RH} \geq & \left[\frac{\Omega^{\rm (0)}_{\rm R}\, h^2}{5.61\times 10^{-6}\,\Delta N_{\rm eff}}\,
\frac{H_{\rm end}^2}{12\, \pi^2\, {M_{\rm P}}^2}\, \frac{\mu(w_{\rm RH})\,(1+3\,w_{\rm RH})}{2\,\pi\,
(3\,w_{\rm RH} -1)}\,\right]^{\frac{3\,(1+w_{\rm RH})}{4\,(3\,w_{\rm RH} -1)}}\, \nonumber \\
 & \times \left(\frac{90\,H_{\rm end}^2\,{M_{\rm p}}^2}{\pi^2\,g_{*\mathrm{RH}}}\right)^{\frac{1}{4}} \equiv T_{\rm RH}^{\rm GW}.
\label{eq:BBNrestriction}
\end{eqnarray}
$\mu_{\rm RH}$ is an $\mathcal{O}(1)$ parameter (for the exact form, see Refs.~\cite{Haque:2021dha,Haque:2025uri}). Setting $T_{\rm RH}^{\rm GW} \sim T_{\rm BBN} \approx 4 \, {\rm MeV}$~\footnote{By incorporating galaxy survey data alongside BBN and CMB observations, the lower bound on the reheating temperature get modified as pointed out in~\cite{Barbieri:2025moq,deSalas:2015glj}.}, it follows that the $\Delta N_{\rm eff}$ constraint on the PGW spectrum becomes impactful only when $w_{\rm RH} > 0.58$ (see, Table.~\ref{GWtre}). This leads to a novel lower bound on the reheating temperature, denoted by $T_{\rm RH}^{\rm GW}$, arising from the PGWs. This bound has important implications for constraining specific inflationary models, as we will explore in the next section.

\begin{figure*}[!ht]
\centering
\includegraphics[width=0015.0cm,height=05.2cm]{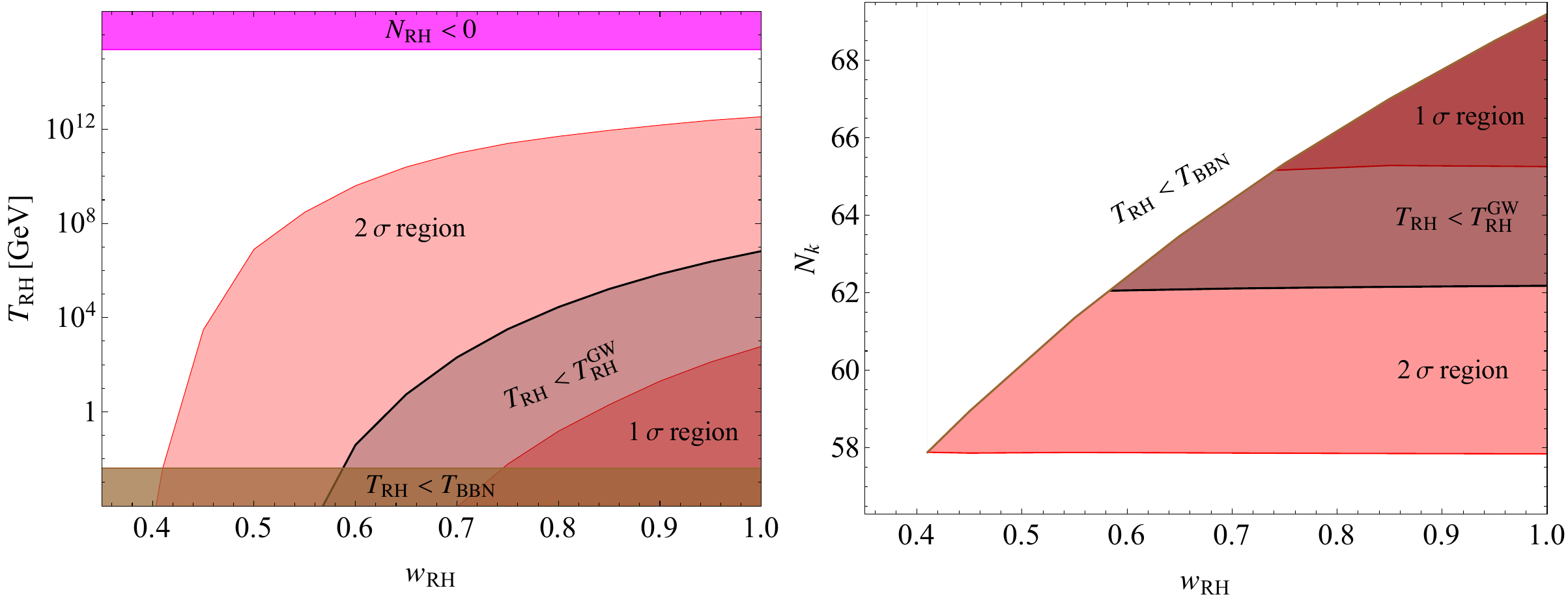}
\caption[]{\justifying \em \textbf{Left panel:} Constraints on the reheating temperature as a function of the reheating EoS $w_{\rm RH}$, based on the $1\sigma$ and $2\sigma$ confidence intervals from the recent \text{P-ACT-LB-BK18} data, are shown as dark and light red regions, respectively. The magenta band marks the region where $N_{\rm RH} < 0$, corresponding to unphysical reheating scenarios, while the brown band indicates temperatures below the BBN threshold, $T_{\rm RH} < T_{\rm BBN}$.The black shaded region indicates the restriction from the $\Delta N_{\rm eff}$ constraints accounting for PGWs.
\textbf{Right panel:} Variation of the number of inflationary e-folds, $N_k$, as a function of the reheating EoS. Other descriptions of this figure are the same as left panel.
}
\label{fig:trh_nk}
\end{figure*}

\section{Results and Discussions}
\label{sec:results}
In this section, we present updated constraints on the Higgs-Starobinsky inflation and its associated reheating dynamics using the current P-ACT-LB-BK18 dataset, together with limits on the overproduction of primary gravitational waves (GWs) inferred from the effective number of relativistic species, $\Delta N_{\rm eff}$, during BBN. As previously discussed, tensor perturbations generated from vacuum fluctuations during inflation inevitably produce PGWs, which must be accounted for any consistent analysis. The post-inflationary reheating phase significantly affects the resulting stochastic GW background and, for a very stiff EoS, can lead to an overproduction of GWs.
A key motivation for our analysis is that, upon incorporating the $\Delta N_{\rm eff}$ bound from PGWs, the entire parameter space allowed at the $1\sigma$ C.L. for the P-ACT-LB-BK18 dataset as shown in Refs.~\cite{Drees:2025ngb,Zharov:2025evb,Liu:2025qca} is effectively ruled out. We find that only a viable region at the $2\sigma$ C.L. Note that throughout this work, BBN constraints refer to the lower bound on the reheating temperature, specifically $T_{\rm BBN} \approx 4\,\mathrm{MeV}$~\cite{Kawasaki:1999na,Kawasaki:2000en,Hasegawa:2019jsa}.

\begin{table*}[!ht]
    \centering
    \renewcommand{\arraystretch}{1.2}
    \begin{tabular}{c|c|c|c|c|c|c}
    \hline
    \hline
        \multirow{2}{*}{C.L.} & \multicolumn{2}{|c}{range of $w_{\rm RH}$} & \multicolumn{2}{|c}{range of $T_{\rm RH}$~[GeV]} & \multicolumn{2}{|c}{range of $N_k$}\\
        \cline{2-7}
        & BBN & BBN+PGW & BBN & BBN+PGW & BBN & BBN+PGW \\
        \hline
        $1\sigma$ & $[0.75-1.0]$ & $-$ & $[T_{\rm BBN}-6.0\times 10^2]$ & $-$ & $[65.7-69.2]$ & $-$ \\
        $2\sigma$ & $[0.41-1.0]$ & $[0.41-1.0]$ & $[T_{\rm BBN}-3.4\times 10^{12}]$ & $[T_{\rm BBN}-3.4\times 10^{12}]$ & $[57.9-69.2]$ & $[57.9-62.2]$ \\
        
    \hline
    \hline
    \end{tabular}
    \caption[]{\justifying \it Comparing the constraints on $w_{\rm RH}, T_{\rm RH}, N_k$ inferred from Higgs-Starobinsky model with post-inflationary reheating dynamics from latest {\rm P-ACT-LB-BK18} dataset, considering constraints from both BBN only and BBN+PGW.}
    \label{tab:model_constraint}
\end{table*}

Fig.~\ref{fig:nsr_model} illustrates the predictions of the Higgs–Starobinsky model in the $(n_s, r)$ plane for various values of $w_{\rm RH}$. For $w_{\rm RH} < 1/3$, the lower bound on $n_s$ is set by the BBN energy scale, while the upper bound corresponds to the maximum reheating temperature allowed by instantaneous reheating. Conversely, for $w_{\rm RH} > 1/3$, this trend is reversed. Our analysis shows that the Higgs–Starobinsky model enters the $2\sigma$ C.L. region of the current $(n_s, r)$ constraints once $w_{\rm RH} \gtrsim 0.41$. Moreover, for $w_{\rm RH} \gtrsim 0.75$, the model prediction falls within the $1\sigma$ C.L., {\it{provided only the BBN constraint is considered}}. However, upon including the PGW constraint alongside the BBN restriction on the reheating temperature, we find that all allowed values of $w_{\rm RH}$ are excluded at the $1\sigma$ C.L.

In Fig.~\ref{fig:trh_nk}, we present the constraints on the reheating temperature, $T_{\rm RH}$, and the number of inflationary e-folds, $N_k$, as functions of the reheating EoS parameter, $w_{\rm RH}$. The $2\sigma$ and $1\sigma$ upper bounds on $T_{\rm RH}$ represent the maximum reheating temperatures at which the model predictions remain within the respective confidence intervals. For $0.41 \lesssim w_{\rm RH} \lesssim 0.58$, the lower bounds on $T_{\rm RH}$ are set by the BBN constraint. However, for $w_{\rm RH} \gtrsim 0.58$, the constraints from PGW overproduction become increasingly stringent, pushing the lower bound on $T_{\rm RH}$ well above the BBN threshold. As a result, the entire $1\sigma$ region is ruled out due to PGW overproduction, that we can clearly see in Fig.~~\ref{fig:trh_nk}. One interesting observation is that, as $w_{\rm RH}$ increases beyond 0.41, the allowed ranges for both the reheating temperature $T_{\rm RH}$ and the number of inflationary e-folds $N_k$ also increase, up to approximately $w_{\rm RH} \sim 0.58$. Beyond this point, the PGW constraint becomes dominant, significantly shrinking the allowed parameter space. For example, at the $2\sigma$ C.L., the allowed range for $N_k$, considering only the BBN bound on the reheating temperature, spans from $(57.9-69.2)$. However, once the PGW constraint is included, this range narrows to $(57.9–62.2)$. 
The detailed numerical results are summarized in Table~~\ref{tab:model_constraint}, comparing both the BBN-only and BBN+PGW bounds.

Therefore, our analysis demonstrates that the Higgs-Starobinsky model remains consistent with current observational data only at the $2\sigma$ C.L. when both BBN and PGW constraints are taken into account. The inclusion of PGW bounds significantly tightens the allowed ranges for both the reheating temperature and the number of inflationary e-folds.

\section{Conclusion}
\label{sec:conclusion}

In this letter, we present updated constraints on the Higgs–Starobinsky inflation model with post-inflationary reheating dynamics, utilizing the latest P-ACT-LB-BK18 dataset. Our analysis incorporates both the BBN constraint on the reheating temperature and the bounds on PGWs overproduction, derived from limits on the effective number of relativistic species, $\Delta N_{\rm eff}$. We emphasize the critical role played by the PGW overproduction bound, which arises from tensor perturbations generated by vacuum fluctuations during inflation, without considering any additional source term. Our key findings are summarized as follows:
\begin{itemize}
\item  In this analysis the reheating equation-of-state parameter is treated as a free variable in the range $0 \leq w_{\rm RH} \leq 1$. However, the Higgs-Starobinsky model is disfavoured by the latest P-ACT-LB-BK18 data at the $2\sigma$ level for $w_{\rm RH} < 0.41$. Thus, the whole analysis is important for $w_{\rm RH}\geq 0.41$, where the model is allowed at $2\sigma$ level. 
\item The novelty of our analysis lies in incorporating the $\Delta N_{\rm eff}$ bound from PGWs, which becomes relevant only for blue-tilted spectra (i.e. $w_{\rm RH} > 1/3$).
Incorporating the PGW constraint in addition to the BBN bound reduces the allowed range of inflationary e-folds at the \(2\sigma\) C.L. from \((57.9\text{–}69.2)\) to \((57.9\text{–}62.2)\). At the \(1\sigma\) C.L., the entire parameter space is ruled out due to PGW overproduction. A similar trend is observed for the reheating temperature: while the BBN bound sets the initial lower limit, increasing the equation-of-state parameter \(w_{\rm RH} > 0.58\) causes the PGW constraint to further restrict the allowed reheating temperature. A stiffer post-inflationary equation of state leads to a more severe reduction in the allowed temperature range (see, for instance, the \textit{left panel} of Fig.~\ref{fig:trh_nk}).
\end{itemize}
In summary, constraints from PGWs play a pivotal role in testing the viability of inflationary models, as demonstrated here for Higgs-Starobinsky inflation. Notably, the PGW-induced bound on $\Delta N_{\rm eff}$ excludes the Higgs-Starobinsky model at the $1\sigma$ confidence level when considering the latest P-ACT-LB-BK18 dataset, despite its otherwise consistent predictions shown in Refs.~\cite{Drees:2025ngb,Zharov:2025evb,Liu:2025qca}. This result imposes additional restrictions on post-inflationary dynamics, particularly on the reheating phase. A more comprehensive Markov Chain Monte Carlo (MCMC) analysis incorporating the latest ACT data could further tighten constraints on reheating parameters and inflationary observables—a direction we leave for future work.

\section*{Acknowledgements}
The Authors gratefully acknowledge the use of computational facilities at  ISI, Kolkata. MRH acknowledges ISI Kolkata for providing financial support through Research Associateship. SP thanks CSIR for financial support through Senior Research Fellowship (File no. 09/093(0195)/2020-EMR-I). DP thanks ISI Kolkata for financial support through Senior Research Fellowship.

\bibliographystyle{elsarticle-harv} 
\bibliography{biblio}

\end{document}